\documentclass[a4paper]{jpsj2}

\title{Quasi-One-Dimensional Spin-Density-Wave States with Two Kinds of 
  Periodic Potentials and a Interchain Misfit}
\usepackage{graphicx}
\usepackage{dcolumn}
\usepackage{bm}
\def\ggs{\buildrel\textstyle > \over {\hbox{\raise0.2ex\hbox{$\sim$}}}}
\def\lls{\buildrel\textstyle < \over {\hbox{\raise0.2ex\hbox{$\sim$}}}}
\def\gsim{\,\lower0.75ex\hbox{$\ggs$}\,}
\def\lsim{\,\lower0.75ex\hbox{$\lls$}\,}

\def \virg{\;\;,}
\def \point{\;\;.}
\def \d{{\rm d}}

\def \exp{{\rm exp}}
\def \kf{k_{\rm F}}

\title{   
Quasi-one-dimensional spin-density-wave states in the presence of 
  two commensurate potentials and an interchain misfit
}

\author
{
T. {\sc Itakura}
\thanks{E-mail address: toshifumiitakura@scphys.kyoto-u.ac.jp}
}

\inst{Department of Physics, 
 Kyoto-University,
 Kyoto 606-8502, Japan \\
 TEL:075-753-3750 \\
} 
\recdate{\today}

\abst
{
 Spin density wave (SDW) states of a quasi-one-dimensional system 
  with an incommensurate wave vector   perpendicular to the chain 
 have been studied  in the presence of 
 two kinds of commensurate potentials, which originate in 
   a quarter-filled band and dimerization along the chain. 
 In terms of a phase variable of the SDW order parameter,  
 we treat classically the two-dimensional Hamiltonian,
    which includes both acoustic excitations with long wave length 
     and a vortex excitation with short wave length. 
  A phase diagram  
  on the plane of temperature and chemical potential 
   (where the latter corresponds to 
     the deviation of the transverse wave vector 
       from the commensurate one)
         exhibits  a variety of states given by 
  the commensurate SDW state without charge density, 
  the commensurate SDW state with charge density, 
  the incommensurate SDW state and the disordered state.   }

\kword
{
spin density wave, phase, 
quarter-filled band,
commensurate-incommensurate transition,
renormalization group method, vortex,  dimerization
  }

\begin{document}
\sloppy
\maketitle
\section{Introduction}
        


In several kinds of density wave systems, 
  a  commensurate-incommensurate transition
 has been studied 
   theoretically
     by varying a misfit  from the commensurate wave vector   
      (or  a chemical potential). 
  Such a transition is followed  by 
    a  formation of the soliton of the phase  of 
    density wave, 
      since the energy gain of  
         a commensurability  competes with 
          the increase of elastic energy driven by  the misfit. 
 Pokrovsky and Talapov have first studied such a phenomenon in  
   a two-dimensional classical system  by using  
    a renormalization group method.
     \cite{Pokrovsky_Talapov_PRL}
 Such a method has been applied to 
     the sine-Gordon model\cite{Horovitz,Tsuchiizu}
    and the Coulomb gas model
    \cite{Itakura_Suzumura_JPSJ01}. 

In contrast to the one-dimensional case,   
   it is crucial to take into account  the  effect of 
   a vortex  in the two-dimensional  case,  which 
 appears at finite temperatures   due to  a thermal  excitation.     
    \cite{Kosterlitz_Thouless}
   The vortex  destroys 
      the quasi-long range order 
       resulting in  the disordered phase.
  The phase diagram of commensurate and incommensurate states 
    in the presence of the vortex interaction  has been studied  
   by using  the renormalization group method (for $\mu=0$) 
   \cite{Jose,Boyanovsky}
   and  the exact result (for finite $\mu$),
   \cite{Haldane,Schulz_PRL} 
 where $\mu$ is the chemical potential corresponding to  the misfit. 
 The commensurability potential is characterized by 
 the $p$-fold periodic potential with respect to 
  the phase.   
 The case for $p = 2$ 
   has been studied extensively    by applying 
   the  renormalization group method  to 
    a  one-dimensional quantum system,
      \cite{Giamarchi_Schulz}        
  in which  the  vortex and  the misfit
    correspond to  the anisotropic interaction of the spin freedom
    and an external magnetic field, respectively.

These studies have been devoted to the  system only with 
   a single  commensurate potential  
 while that with  two kinds of commensurate potentials 
 is of interest to investigate 
  the following   quasi-one-dimensional conductors, Bechgaard  salts. 
\cite{Jerome,Yamaji,Bechgaard} 
 The salts, 
  which  have a dimerized and quarter-filled conduction band,   
 exhibit   the spin density wave (SDW) state at low temperatures 
 and undergoes a  commensurate-incommensurate transition  
  perpendicular to chain    under effective pressures.  
 \cite{Moser}
  The interchain interaction  including     
   the misfit of  the  phase difference between chains 
    competes with intrachain commensurability energies  
     where the misfit originates in 
      the nesting of Fermi surface
\cite{Yamaji_JPSJ} 
 and/or  exchange energy due to the various kinds of interchain hopping.
\cite{Ducasse,Itakura_Suzumura_JPSJ01}    
 When  the repulsive interaction for the nearest-neighbor sites 
 is large,   the commensurability energy at quarter-filling 
   leads to the coexistence of SDW with  the charge ordering. 
   \cite{Seo,Suzumura_JPSJ97}
 However, the 
   dimerization leads to  an effectively 
   half-filled band and is in favor of 
 pure SDW state (i.e., the absence of charge ordering).  
 In terms of  a transfer integral method, 
  the role of interchain misfit in the  system  with  
    these two kinds of  potentials has  been examined    
   for    two-coupled chains 
    \cite{Itakura_Suzumura_JPCS}      
 and for a quasi-one-dimensional system, 
 where  the phase diagram for the latter model has been obtained 
 on the plane of  an intrachain misfit  and temperature. 
   \cite{Itakura_Suzumura}

 In the present study, 
   we examine the effect of these two commensurate energies 
   on  the quasi-one-dimensional   quarter-filled SDW system  
    with a transverse misfit  
   by adding   dimerization and vortex to the previous model.
  \cite{Itakura_Suzumura_JPSJ01}
 In \S 2, 
    the classical two dimensional Hamiltonian for SDW state is derived 
  by expanding   the interchain coupling   up to the second order 
   with respect to phase difference between chains. 
   The renormalization group equations   
       are derived. 
  In \S 3, we obtain  the phase diagram  of the commensurate SDW state and 
   the incommensurate SDW state and the disordered state 
 (induced by the  vortex). 
 In \S 4, discussion is given.

\section{Formulation}
 We consider the SDW state of the quasi-one-dimensional electron system 
  with the mean-field order parameter given by  
     $\Delta_{2\kf}^i \cos(2\kf x_0+\theta_i)$ 
 where $\Delta_{2\kf}^i$ and $\theta_i$ 
   denote the amplitude and the phase
  of the $i$-th chain, 
 respectively.  
   The effective Hamiltonian 
   describing   the charge fluctuation in the classical case   
    is written as  
\begin{eqnarray}
   {\cal H}_{\rm Q1D} = 
  \sum_{i} \int {\rm d} x_0  
    \biggl[ A \left( \frac{ \d \theta_i(x_0)}{ \d x_0 } \right)^2 
    - W_c \cos 4 \theta_i(x_0) 
\nonumber \\ &&{}
     -W_d \sin 2 \theta_i (x_0)
\nonumber \\ {}
   +J_0 \cos ( \theta_i(x_0) -  \theta_{i+1}(x_0)  +  \mu )  
       \biggr] ,
   \label{eqn:KT}
\end{eqnarray}
where the $A$-term is an elastic energy while 
 the $W_c$- and $W_d$-terms correspond to 
  the  commensurability energy 
   at quarter-filling and 
 that with  dimerization, respectively.   
Equation  (\ref{eqn:KT}) for two-chain has been studied previously.  
   \cite{Itakura_Suzumura_JPCS} 
In the case of $W_d=J_0=0$,
   the spatial variation with a periodic array of  
   $(\uparrow, 0, \downarrow, 0)$ is obtained 
 for    $W_c>0$, which is found  for 
   large nearest-neighbor repulsion.  
\cite{Seo,Suzumura_JPSJ97}  
The $J_0$-term 
    represents the interchain interaction with
    $\mu$ being
    a misfit parameter corresponding to the phase difference  between  
      nearest-neighbor chains.
We examine only the case of $J_0 <0$,
   because the replacement of   $\theta_i \to \theta_i +\pi$ 
   in  the even chain leads to the same result as that of  $J_0>0$.
Since the calculation of eq.~(\ref{eqn:KT}) is very complicated, 
 we replace it by an effective  one, which consists of 
 acoustic excitation  with long wave length and the excitation 
 with short wave length.   
The former, which  leads to   
 the {\it anisotropic   two-dimensional} SDW state at low temperatures, 
 is obtained by 
  expanding  the $J_0$-term in eq. (\ref{eqn:KT}) with respect to 
   $ \theta_i(x_0) -  \theta_{i+1}(x_0) +  \mu$  
    up to the second order.  The  continuum limit is taken as 
    $\theta_i(x_0)\to \theta(x_0,y_0)$ with $y_0=ib$ where 
    $b$ is the lattice constant for interchain direction.  
 The latter,     
 which corresponds to  the  higher order terms 
 and is dominant at high temperatures near the critical temperature,
     can be represented in terms of  
   the vortex field,\cite{Jose,Boyanovsky}
 which is relevant to 
  the potential for   $\theta$  
    with a  periodicity, $2 \pi$.  
 The potential for the vortex field is expressed as 
\begin{eqnarray}
W_v \cos 2 \pi K \phi (x,y)
\virg
\label{eqn:vortex}
\end{eqnarray}
 which resembles  a  one-dimensional quantum system
 (see Appendix).  
In eq.(\ref{eqn:vortex}),  
   $\phi$ is   the conjugate phase variable  defined as
   $\partial \theta / \partial x_0=i\beta\partial \phi  / \partial y_0$ 
  and
   $\partial \theta / \partial y_0 
  = -i\beta^{-1} \partial \phi  / \partial x_0$ 
  with $\beta = b (|J_0|/2 A)^{1/2}$. 
The quantity $K$ is given by $K\equiv \sqrt{2A|J_0|}/T_0$,
and the quantity $W_v$ is the coupling constant for the vortex field
    whose form depends on a microscopic model.
 The quantity $T_0$ is the temperature.  
 By use of  
 dimensionless quantities, 
\begin{eqnarray*}
x &=& \sqrt{\frac{W_c}{A}}\, x_0, \,\,
y = \frac{y_0}{b},\,\,
T =\frac{T_0}{\sqrt{AW_c}} , \,\,
J =\frac{|J_0|}{W_c} ,
\end{eqnarray*}
 the effective Hamiltonian for eq.(\ref{eqn:KT})
   is given by \cite{Itakura_Suzumura_JPSJ01}
\begin{eqnarray}
     \label{eqn:KTV}
\frac{ {\cal H}_{\rm Q1D}}{T_0} 
&=& 
\int \int \d x \d y 
\nonumber \\ &\times &
\biggl[ 
  \frac{1}{T} 
\left( \frac{\partial \theta (x,y)}{\partial x} \right)^2 
  + 
 \frac{J}{2T}  
  \left( \frac{\partial \theta (x,y)}{\partial y} - \mu \right)^2  
\nonumber \\ && {} 
  - \frac{1}{T} \cos 4 \theta (x,y) 
  - \frac{1}{T}\left(\frac{W_d}{W_c}\right) \sin 2 \theta (x,y)
\nonumber \\ && {} 
  - \frac{1}{T}\left(\frac{W_v}{W_c}\right) \cos 2 \pi K \phi(x,y)
          \biggr] 
 \point
\end{eqnarray}
For the convenience of applying a renormalization
   group method, 
  eq.~(\ref{eqn:KTV}) is rewritten as 
\begin{eqnarray}
\frac{ {\cal H}_{\rm Q1D}}{T_0} 
&=& 
\int \int \d x \d y 
\nonumber \\ &\times &
\biggl[ 
  \frac{K}{2\beta} 
\left( \frac{\partial \theta (x,y)}{\partial x} \right)^2 
  + 
 \frac{K \beta}{2}  
  \left( \frac{\partial \theta (x,y)}{\partial y} - \mu \right)^2  
\nonumber \\ && {} 
  - 2 y_4 \cos 4 \theta (x,y) 
  - 2 y_2 \sin 2 \theta (x,y)
\nonumber \\ && {} 
  - 2 y_v \cos 2 \pi K \phi(x,y)
          \biggr] 
 \virg
\label{eqn:PHSTRG}
\end{eqnarray}
  where
   $K=\sqrt{2J}/T$, $\beta\equiv\sqrt{J/2}$, $y_4=1/2T$,
   $y_2\equiv(W_d/W_c)/2T$ and $y_v\equiv(W_v/W_c)/2T$.
The renormalization group equations are calculated 
    by use of Coulomb gas method while  
 the same results are obtained 
     by the use of the  method of  Giamarchi and Schulz.
    \cite{Giamarchi_Schulz}
These    equations are obtained   as 
\begin{eqnarray}    
  \label{eqn:rgk}
 \frac{ \d }{\d l}\frac{1}{K} 
 &=&  -\frac{4 \pi \overline{y}_2^2}{K^2} J_0 ( 2 \overline{\mu}) 
    -\frac{16 \pi  \overline{y}_4^2}{K^2} J_0 ( 4 \overline{\mu}  ) 
    + 4 \pi^3 \overline{y}_v^2
 \virg  \\
\label{eqn:rgy2}
\frac{ \d \overline{y}_2}{\d l} 
&=& 
\left( 2 - \frac{1}{ \pi K} \right) \overline{y}_2 
 - 2 \pi \overline{y}_2 \overline{y}_4 
\virg \\
\label{eqn:rgy4}
\frac{ \d \overline{y}_4}{\d l} 
&=& 
\left( 2 - \frac{4}{ \pi K}\right) \overline{y}_4 
 - \pi \overline{y}_2^2 
\virg \\
\label{eqn:rgyv}
\frac{ \d \overline{y}_v}{\d l} 
&=& 
\left( 2 - \pi K \right) \overline{y}_v
\virg \\
\label{eqn:rgm}
\frac{\d \overline{\mu}}{\d l} 
=
\overline{\mu} &-& 
   \frac{ 4 \pi \overline{y}_2^2    }{K } J_1 (2 \overline{\mu} ) 
   - \frac{ 8 \pi \overline{y}_4^2    }{ K } J_1 (4 \overline{\mu}  )  
\virg   \\
\label{eqn:rgb}        
\frac{ \d \ln \beta}{\d l} 
&=&
            \frac{4 \pi \overline{y}_2^2}{K} J_2 ( 2  \overline{\mu} )
            + \frac{16 \pi   
            \overline{y}_4^2}{ K} J_2 ( 4 \overline{\mu}) 
\end{eqnarray}
 where  
 $\overline{y_4}\equiv \beta \alpha^2 y_4$,
 $\overline{y_2} \equiv \beta \alpha^2 y_2$, 
 $\overline{y_v}\equiv \beta \alpha^2 y_v $
   and $\overline{\mu}\equiv \beta \alpha \mu $.
The quantity  $\alpha$
   is short-length cutoff of the order of lattice constant
 where the constant  is taken as unity.
We note that the second order corrections 
 to eqs.~(\ref{eqn:rgy2}) and (\ref{eqn:rgy4})  
    are obtained by   examining 
   the response function up to third order,  $y_2^2 y_4 $. 
  \cite{Tsuchiizu_Dr,Tsuchiizu_pre}

\section{Phase Diagram}
 
 The renormalization group  equations
 (eqs.~(\ref{eqn:rgk})-(\ref{eqn:rgm})) are calculated  numerically
 by choosing initial conditions as  
    $1/K (=T/(2J)^{1/2})$,  
   $\overline{\mu} (= \alpha \mu (J/2)^{1/2}))$, 
  $y_4 = \alpha^2 K /4$ and  $y_2 =(W_d/W_c) y_4$. 
 The vortex interaction term is taken as 
   $\overline{y}_v= \exp( - \pi^2 K /2 )$
  as found in  the isotopic XY model.
 \cite{Jose}
 In the present calculation, we take 
 $\alpha=0.5$  and 
  use an approximation for  the Bessel function 
 such that $J_0(z) = 1$, $J_1(z)=z/2$ for $z \lsim 1.8$ 
 and zero otherwise.   
These are  possible choices to reproduce the well known result
 for $y_2=0$ (see Fig. 1).  

From the solution of the renormalization equations, we obtain   
  the  following states. 
(i)The irrelevant  $\overline{\mu}$ leads to 
 two kinds of  commensurate 
 states.   The relevant $y_2$ 
 corresponds to the commensurate state of pure SDW (C$_{\rm I}$-phase)
 while the irrelevant $\overline{y_2}$ 
 and relevant $\overline{y_4}$ correspond to 
  the commensurate SDW state coexisting with charge order (C$_{\rm II}$-phase)
(ii) When   $\overline{\mu}$ is relevant,  
  the incommensurate state (IC-phase) is obtained for 
 the irrelevant $\overline{y_v}$ and 
 the disordered state with the vortex (disorder-phase) 
 is obtained for    the relevant $\overline{y_v}$. 
These fixed points are summarized in Table I where $K^*$ 
 denotes a finite value.     
\begin{table}[t]
\begin{tabular}{@{\hspace{\tabcolsep}\extracolsep{\fill}}cccccc}
    \hline
     & $1/K$ & $\overline{y_2}$ & $\overline{y_4}$
        & $\overline{y_v}$ & $\overline{\mu}$ \\ 
 \hline 
 C$_{\rm I}$ &  $0$ & $\infty $  & $ - \infty$ &  0   & 0     \\ 
 C$_{\rm II}$ & $0$ &  0  &  $\infty$   &   0  & 0   \\ 
 IC  &  $1/K^*$    & (0)         & (0)  
 &   0  & $\infty$ \\ 
 Disorder  & $\infty$   &  0 
           & 0        &  $\infty$  & $\infty$ \\ \hline
\end{tabular} 
\caption{
Fixed point values for respective phases.
The parenthesis indicates the expected value.
}
\end{table}

First, we examine the case that the dimerization is absent ($y_2=0$). 
  In the present calculation, we take 
 $\alpha=0.5$  and 
  use an approximation for  the Bessel function 
 such that $J_0(z) = 1$, $J_1(z)=z/2$ for $z \lsim 1.8$ 
 and zero otherwise.   
These are  possible choices, which  reproduce  well 
  the previous results.
\cite{Haldane,Schulz_PRL}
For  small $\overline{\mu}$, the smaller $\alpha$ enhances the IC region 
 while the larger $\alpha$ diminishes the IC region 
In Fig. 1, there are three fixed points corresponding to 
 C$_{\rm II}$, IC and disorder phases (Fig. 1).
In the C$_{\rm II}$-phase,  the charge density becomes  long range 
while,  in the IC-phase, the gapless  excitation  due to 
 a finite $K^* (= K(\infty))$  
   results in  the power low decay of the response function
   of density waves.
   \cite{Itakura_Suzumura_JPSJ01}
In  the disorder  phase, 
 the vortex, which is activated by thermal fluctuations,
   destroys the (quasi-)long range order and
 the  response function for the density wave decays exponentially. 
    There exists the  incommensurate state between
    the commensurate state and the disordered state 
    if  $\overline{\mu} \ne 0$. 
The absence of the incommensurate state
    for $\overline{\mu}$=0 is understood as follows.  
 Equations (\ref{eqn:rgy4}) and (\ref{eqn:rgyv})  shows 
  the irrelevant  $\overline{y_4}$  ($\overline{y_v}$)  
      for $1/K^* > \pi/2 $ ($1/K^* < \pi/2$)
    so that either $\overline{y_4}$ or $\overline{y_v}$ is relevant for 
    $\overline{\mu}$=0.
We remark  followings for other boundaries.
Equation (\ref{eqn:rgyv}) shows that
 $1/K^* $ is $\pi/2$ at the boundary between  disorder phase and
 IC-phase.
The boundary between  C$_{\rm II}$ and IC at T=0 (i.e., $1/K = 0$)
  is given by $\beta \mu = 2 \sqrt{2}/\pi$, which can be  derived 
  analytically by examining the condition for 
  the formation of the soliton.   
For $\beta \mu >  2 \sqrt{2}/\pi$,
   $\overline{y_4}$ is always irrelevant,
   and the critical temperature for $\overline{y_v} (l) \rightarrow 0$
   is almost independent of $\overline{y_4}$.
The boundary at $1/K=0$ is obtained from 
  eqs. (\ref{eqn:rgy2}) and (\ref{eqn:rgyv}),
 which lead to  
  $ \d (\overline{y_4}^2 - 2 \overline{y_2}^2) / \d l 
  = 4  (\overline{y_4}^2 - 2 \overline{y_2}^2 )$.
When temperature is decreased 
 (i.e., $\overline{y}_4$ is increased) with  
  fixed $y_2/y_4 = W_d/W_c$, it turns out that 
 the C$_{\rm I}$ phase moves into the C$_{\rm II}$ phase 
  for  $y_2 < \sqrt{2} y_4$ corresponding to the small dimerization. 
With increasing $1/K$,
   the region of C$_{\rm I}$-phase increases,
   because the effect of $1/K$ on the  scaling dimension
   for $\overline{y_4}$  (i.e., $2 -4/\pi K$)
    is larger than 
     that for $\overline{y_2}$ (i.e., $2 -1/\pi K$).
 In addition,   
 the high temperature corresponds to large $1/K$ due to 
 $1/K = T/\sqrt{2 J}$ so that  
  C$_{\rm I}$ phase (C$_{\rm II}$ phase ) 
   is obtained at high (low) temperatures.  
With increasing $1/K$ from  the C$_{\rm II}$-phase which is
 obtained for  $1/K^* < \pi/2$, 
 one always finds  the   C$_{\rm I}$-phase, which is located 
 in    $\pi/2 <1/K^* < 2 \pi$. 

Next we examine the state in the presence of 
 both dimerization and quarter-filled commensurability energy,
  which  exhibit  four fixed points as shown in Table 1. 
 Their states are given by 
 two kinds of commensurate states. 
  (C$_{\rm I}$-phase and C$_{\rm II}$-phase),
   incommensurate state (IC-phase) and disordered state
      (disorder-phase). 
In Fig.3, the phase diagram is shown on
  the plane of $\beta \mu$ and $1/K$ 
   where  $y_2/y_4 = W_d/W_c = 0.5$.   
The C$_{\rm I}$-phase (the commensurate state
   of the  pure 2$k_F$ SDW state) is found  at high temperatures 
 and for $\beta \mu$ smaller than a critical value.   
  The C$_{\rm II}$-phase (the commensurate state 
   coexisting  with 4$k_F$ CDW state) is found at low temperatures
  and the critical value of $\beta \mu$ is larger than  
  that of  C$_{\rm I}$-phase.   
  Compared with Fig. 1, 
   it turns out that dimerization has an effect of 
     producing the region of  C$_{\rm I}$-phase around 
    the triple point in Fig. 1,   at which  
            disorder, IC and C$_{\rm II}$ merge together.
 Note that the boundary between C$_{\rm II}$ and IC at $T (=1/K)=0$ 
 does not change within the visible scale even in the presence of both 
 $y_2$ and $y_v$, which become irrelevant at low temperatures.  
With decreasing temperature with fixed $\beta \mu$, 
  one obtains following  four kinds of successive transitions: 
(I)  disorder $\to$ C$_{\rm I}$ $\to$ C$_{\rm II}$,
(II)  disorder $\to$ C$_{\rm I}$ $\to$  IC $\to$ C$_{\rm II}$,
 (III) disorder $\to$ IC $\to$ C$_{\rm II}$, and  
(IV)disorder $\to$ IC.
 However the tricritical point for C$_{\rm I}$ C$_{\rm II}$ and IC 
  depends  on the choice of $\alpha$ 
  which leads to   another transition of   
    disorder $\to$  IC $\to$ C$_{\rm I}$ $\to$ C$_{\rm II}$ 
   instead of (II). 
 Therefore the analysis of 
  such a detail is  beyond the present scheme of 
 the renormalization group equation. 
   The remained three kinds of transition  
      seem to be reasonable  
  as the result of combining  the commensurability of 
 $p=2$   and  that of  $p=4$. 

 
\section{Discussion} 
We examined the effect of transverse misfit (chemical potential) 
 on the quasi-one-dimensional 
 SDW system which includes two-kinds of commensurate potentials 
 coming from the dimerized quarter-filled band. 
The anisotropic two-dimensional SDW system is treated classically  
 since    
  the  quantum fluctuations may have an effect of 
 renormalizing the amplitude of $W_c$ and $W_d$ but does not change 
 qualitatively the form of the potential.  
We found 
  the  phase diagram, which exhibits a variety of states,
  on the plane of chemical potential 
  and temperature   
 when the two kinds of commensurate potentials 
 compete with each other. 
The transition from the disordered state 
  to the incommensurate SDW state 
  is the Kosterlitz-Thouless transition.
It is of interest whether such a  transition 
   in addition to commensurate-incommensurate transition
   is related to the sub-phases of (TMTSF)$_2$PF$_6$ salt.
   \cite{Takahashi_sub}
      
Finally we note  the phase diagram  of  Fig. 3,  
 which is the result obtained 
 in the classical  quasi-one-dimensional system 
  with   two-kinds of periodic potential competing each other, 
   anisotropic misfit parameter, 
   and vortex interaction induced by  interchain coupling. 
 This result in two-dimensional classical case 
 has a common feature with the previous one,
\cite{Itakura_Suzumura} 
  (except for the state around the triple point)  
 in which  
 the interchain coupling is treated in the mean-field theory 
  and  the transfer integral method is applied to treat 
   the thermal fluctuation exactly. 
 In these  cases, 
  two kinds of potentials are the same 
 but   the misfit in the previous  case is  in the chain  and    
 the present  case is perpendicular to the  chain. 
Compared with the phase diagram (Fig. 8 in the previous case
 \cite{Itakura_Suzumura}), 
 we found the similarity   
 in the sense that 
 C$_{\rm I}$ (C$_1$) is the pure SDW state, 
 C$_{\rm II}$ (C$_2$) is the SDW  state coexisting with the CDW  
 and the disorder phase denotes  the absence of the long range order, 
 (i.e., $\langle \cos \theta \rangle = 0$).
 The notable difference  is  between the incommensurate along the chain 
 and that perpendicular to chain.  
 It is considered that such a nice  correspondence  
   between these two results may be attributable to  the 
   proper treatment of the thermal fluctuation.

\section*{ Acknowledgements }
This work is supported by the Grant-in-Aid for the 21st Century COE "Center for Diversity and Universality in Physics" from the Ministry of Education, Culture, Sports, Science and Technology (MEXT) of Japan.
\appendix
\section{Classical 2D Hamiltonian vs.  quantum 1D Hamiltonian}
Using the path integral methods,
 we show that the quantum one-dimensional (1D model 
   is equivalent to the present classical two-dimensional (2D) model.
Using the Stratonovich-Hubbard identity,
\cite{Negele}
\begin{eqnarray}
 \int {\cal D} \theta_i \, \exp &&
 \left[ - \int \d x_0 \sum_i \frac{A}{T_0} 
    \left(\frac{\d \theta_i (x_0)}{\d x_0} \right)^2 \right]  
    \nonumber \\
    &=&
    \int {\cal D} \theta_i  \,{\cal D} \Pi_i \,
  \nonumber \\ && {} \times
    \exp \left[ - \int \d x_0 \sum_i 
    \left\{ \frac{T_0}{4 A} \Pi_i^2 
   - i \Pi_i 
    \frac{\d \theta_i}{\d x_0} \right\}  \right] ,
    \nonumber \\
\end{eqnarray}    
the partition function for the two-dimensional Hamiltonian 
 of eq. (\ref{eqn:KT}) 
is rewritten as 
\begin{eqnarray}
Z &=& \int {\cal D} \theta_i, {\cal D} \Pi_i,
 \nonumber \\  &{}\times& 
 {\rm exp} 
    \biggr[ - \int \d x_0 \sum_i \biggl\{
    \frac{T_0}{4 A} \Pi_i^2  - i \Pi_i  
    \frac{ \d \theta_i}{\d x_0}
    \nonumber \\
 \hspace*{1.5cm} &&
  {} - 
     \frac{W_c}{T_0} \cos 4 \theta_i
   - \frac{W_d}{T_0} \sin 2 \theta_i   
\biggl] \point   
 \nonumber \\  &&
   + \frac{J_0}{T_0} \cos ( \theta_i - \theta_{i+1} + \mu ) 
    \biggr\} \biggr] \point 
\end{eqnarray}
Next, we consider the one-dimensional 
   quantum  Hamiltonian with ($\Delta x$ being the lattice constant) 
given by 
\begin{eqnarray}
 {\cal H}_q &=& \sum_{j} 
 \Delta x   
    \biggl[ \frac{T_0}{4 A}   \Pi_j^2 
\nonumber \\ &&{}
    - \frac{W_c }{ T_0} \cos 4 \theta_j 
     -\frac{W_d }{ T_0} \sin 2 \theta_j 
\nonumber \\ &&{}
 - \frac{g_{1/4}}{2\pi^2\alpha^2} \int \d x \cos 4 \theta_{+}
 - \frac{g_{1/2}}{2\pi^2\alpha^2} \int \d x \sin 2 \theta_{+}
\nonumber \\  \hspace*{1.5cm} {}
   &+& \frac{J_0}{T_0}
    \cos ( \theta_j  -  \theta_{j+1}   +  \mu )  
      \biggr] \virg 
  \nonumber \\  
&&{}+\sum_{i} \int {\rm d}x_0 J_0
   \cos ( \theta_i(x_0) -  \theta_{i+1}(x_0)  +  \delta )  
\virg 
   \label{eqn:QKT}
\end{eqnarray}
   where $ [ \theta_j , \Pi_k  ] 
   =  i \delta_{jk} $ 
 and 
 $j$ denotes the lattice index along $x$-direction.
Starting with eq. (\ref{eqn:QKT}),
the partition function for the Hamiltonian of eq. (\ref{eqn:QKT}) is
\cite{Negele}
\begin{eqnarray}
 Z &=& {\rm Tr} ( e^{- \beta {\cal H}_q} ) 
\nonumber \\
  &=& \lim_{N \to \infty} {\rm Tr}  
    ( e^{- \frac{\beta}{N} {\cal H}_q } )^N
\nonumber \\
  &=& \prod_i \sum_{\theta_i ( \tau_0 ) } \langle \theta_i ( \tau_0 )|
    e^{ - \Delta \tau {\cal H}_q } \prod_j \sum_{\theta_j ( \tau_1 )}
  | \theta_j (\tau_1 )\rangle \langle\theta_j ( \tau_1 ) | 
    \nonumber \\
 \hspace*{2cm} && {} \times e^{- \Delta \tau {\cal H}_q} \cdots
   | \theta_i ( \tau_0 ) \rangle
\nonumber \\
 &=& \int {\cal D} \theta_j  \, {\cal D} \Pi_j 
   \nonumber \\ &&  {}\times 
{\rm exp} \biggl[ - \int_0^{\beta} \d \tau \sum_j 
\Delta x
\nonumber \\
&&  \biggl\{
  \frac{T_0}{ 4 A} \Pi_j^2 ( \tau ) - i  \Pi_j ( \tau ) 
  \frac{\d \theta_j (\tau)}{\d \tau}
  \nonumber \\
\hspace*{1cm} {} 
&-& \frac{W_c}{T_0} \cos 4 \theta_j ( \tau ) 
  - \frac{W_d}{T_0} \sin  2 \theta_j ( \tau ) 
  \nonumber \\
\hspace*{1cm} {}
  &+& \frac{J_0}{T_0} 
  \cos  (\theta_j ( \tau ) - \theta_{j+1} ( \tau ) + \mu)
  \biggr\} \biggr] \virg \nonumber \\
\end{eqnarray}  
where $\Delta \tau = \beta /N$ and $\tau$ is the imaginary time.
Thus, the present classical Hamiltonian of eq. (\ref{eqn:KT})
 is equivalent to
    the  quantum Hamiltonian of eq. (\ref{eqn:QKT}).
By taking   the continuum limit of ${\cal H}_q$ along $x$-direction,
   namely $\theta_j(\tau) \to \theta_+(\tau, x)$ 
   and $\sum_j \to \int \d x /\Delta x $ with $x=j \Delta x$,
   one obtains following effective Hamiltonian,
\begin{eqnarray}
   \label{eqn:phase}
{\cal H} &=& 
 \frac{v_\rho}{4\pi}
\int \d x  
\left[ 
  \frac{1}{K_{\rho}} \left( \partial_x \theta_{+} - \nu \right)^2 
  +
  K_{\rho} \left( \partial_x \theta_{-} \right)^2
\right]
\nonumber  \\
 &+& \frac{g_{1/4}}{2\pi^2\alpha^2} \int \d x \cos 4 \theta_{+}
 - \frac{g_{1/2}}{2\pi^2\alpha^2} \int \d x \sin 2 \theta_{+} 
\virg
\nonumber   \\ 
 &-& \frac{g_v}{2\pi^2\alpha^2} \int \d x \cos \theta_{-} \virg
\end{eqnarray}
with the commutation relation,   
$ [ \theta_{+} (x),\partial_x \theta_{-} (x') ] = - 2 \pi i \delta ( x-x')$
\cite{Suzumura_ph}
and $K_{\rho} = T_0 / ( 2 \pi \sqrt{2A|J_0|} )$, $\nu = - \mu/\Delta x$, 
   $v_{\rho} =  \sqrt{|J_0|/2A}$, 
   $g_{1/4}= -2 ( \pi \alpha )^2 W_c/T_0$
   and $g_{1/2} = 2  ( \pi \alpha)^2 W_d/ T_0 $.
The Hamiltonian eq.(\ref{eqn:phase}) 
 has been examined   
 in the 1-D interacting quarter-filled  electron system with dimerization.
  \cite{Tsuchiizu_pre}

We comment on  the interaction of the vortex 
 in terms of $\theta_-(x)$. 
In eq. (\ref{eqn:QKT}), the Hamiltonian is invariant 
   under the transformation given by  
   $\theta_i  \rightarrow \theta_i  + 2 \pi $.
The excitation expressing  such a spatial variation 
 corresponds to the short range topological excitation which 
  is related to the  vortex excitation.
Actually,  in continuum model,  
  such an  excitation is expressed as follows.  
Using the commutation relation, 
 the vortex excitation is generated by unitary transformation, 
in terms of  $V (x) =e^{ i \theta_- (x)}$,
 that is
 $ V(x') \theta_+ (x) V(x')^{\dagger} = 
 e^{ i \theta_- (x')} \theta_+ (x) e^{-  i \theta_- (x')} =
  \theta_+(x) + \pi {\rm sgn} (x-x')$.
Then taking into account the symmetry of eq. (\ref{eqn:QKT}),
 one can add the potential,
$\int \d x W_v \cos \theta_- (x) $
to eq. (\ref{eqn:phase}),
 since the term  is invariant under the unitary  transformation 
 representing  the lowest vortex excitation 
( the vortex potential which has lowest scaling dimension).
 \cite{Orignac} 
Note that 
  the correspondence  between classical system and quantum system 
 is given by 
     $x \leftrightarrow \tau$, 
     $y \leftrightarrow x$,
     $b \leftrightarrow \Delta x$,
     $\theta \leftrightarrow \theta_{+}$,
     $ 2 \pi K \phi \leftrightarrow \theta_{-}$,
     $K \leftrightarrow 1/2 \pi K_{\rho}$ and 
     $\beta  \leftrightarrow v_{\rho}$.
 Thus   we obtain the effective classical Hamiltonian 
 eq. (\ref{eqn:KTV})
  which includes the vortex excitation. 
However, 
    the microscopic calculation is needed  
     to obtain the coefficient, $W_v$. 
 In the present paper, 
     we put $W_v = {\rm exp}(-\pi^2 K/2 )$, 
 which is the extrapolation  deduced from 
    the isotopic stiffness case.
    \cite{Jose}


\end{document}